\documentclass[conference]{IEEEtran}
\IEEEoverridecommandlockouts
\usepackage{cite}
\usepackage{amsmath,amssymb,amsfonts}
\usepackage{algorithmic}
\usepackage{graphicx}
\usepackage{textcomp}
\usepackage{xcolor}
\usepackage{subfigure}
\usepackage{multirow}
\def\BibTeX{{\rm B\kern-.05em{\sc i\kern-.025em b}\kern-.08em
    T\kern-.1667em\lower.7ex\hbox{E}\kern-.125emX}}
\begin{document}

\title{Performance considerations on execution of large scale workflow applications on cloud functions\\
}

\author{\IEEEauthorblockN{Maciej Pawlik}
\IEEEauthorblockA{\textit{AGH Univ. of Science and Technology} \\
Krakow, Poland \\
m.pawlik@cyfronet.pl}
\and
\IEEEauthorblockN{Kamil Figiela}
\IEEEauthorblockA{\textit{AGH Univ. of Science and Technology} \\
Krakow, Poland \\
kfigiela@agh.edu.pl}
\and
\IEEEauthorblockN{Maciej Malawski}
\IEEEauthorblockA{\textit{AGH Univ. of Science and Technology} \\
Krakow, Poland \\
malawski@agh.edu.pl}
}


\maketitle

\begin{abstract}
Function-as-a-Service is a~novel type of cloud service used for creating distributed applications and utilizing computing resources. Application developer supplies source code of cloud functions, which are small applications or application components, while the service provider is responsible for provisioning the infrastructure, scaling and exposing a~REST style API. This environment seems to be adequate for running scientific workflows, which in recent years, have become an~established paradigm for implementing and preserving complex scientific processes. In this paper, we present work done on evaluating three major FaaS providers (Amazon, Google, IBM) as a~platform for running scientific workflows. The experiments were performed with a~dedicated benchmarking framework, which consisted of instrumented workflow execution engine. The testing load was implemented as a~large scale bag-of-tasks style workflow, where task count reached 5120 running in parallel. The studied parameters include raw performance, efficiency of infrastructure provisioning, overhead introduced by the API and network layers, as well as aspects related to run time accounting. Conclusions include insights into available performance, expressed as raw GFlops values and charts depicting relation of performance to function size. The infrastructure provisioning proved to be governed by parallelism and rate limits, which can be deducted from included charts. The overhead imposed by using a~REST API proved to be a~significant contribution to overall run time of individual tasks, and possibly the whole workflow. The paper ends with pointing out possible future work, which includes building performance models and designing a~dedicated scheduling algorithms for running scientific workflows on FaaS.
\end{abstract}


\section{Introduction}

Scientific workflows are an~established paradigm of implementing and preserving a~scientific process. Workflows allow for modeling complex scientific procedures with help of abstractions over infrastructure or implementation details. The introduced, high level description of the process makes it feasible to provide research reproducibility and reuse parts of the workflow. Scientific workflows are usually executed by a~Scientific Workflow Management System\cite{deelman2009workflows}, which provides features required to automate and streamline the execution. In order to execute the workflow we need two additional components, data to operate on and a~computing infrastructure. The data is usually provided by the scientist or is an~artifact produced by or directly included in the workflow. The infrastructure can be a~personal computer, a~HPC computing cluster or a~cloud infrastructure. Due to the availability, pricing models and possibility to dynamically adapt to the workloads, cloud infrastructure seems to be a~natural choice. One of the newest additions in cloud service provider's portfolios is the Function-as-a-Service. FaaS infrastructures provide computing power while taking the responsibility for on-demand provisioning of execution environments. Additionally FaaS offers an~attractive pricing model where user is billed only for the actual time spent on computing, usually with 100 ms granularity. In the case of such infrastructure the user is responsible only for supplying the application code and declaring memory requirements. Applications destined to run of FaaS are called {\em Serverless Applications} in order to emphasize the lack of traditional servers when we think about the deployment or operation of the application. 

In this paper, we present work done on measuring and analysis of various performance aspects of FaaS. Supplied features, like instant provisioning and high scalability come at the cost of limited function run time and limited resources assigned to individual tasks. In the case of most providers, the function's declared memory quota is proportional to the allocated computing power and to the cost of function run time. The presented results allow to determine available computing power and its accessibility when it comes to standard usage scenarios encountered in scientific workflows. The results can be used to construct a~performance models and determine feasibility of using FaaS for scientific workflows. Presented experiments were conducted on infrastructures supplied by three major FaaS providers: Amazon, Google and IBM.

\begin{figure*}[htb]
\centering
\includegraphics[width=\textwidth]{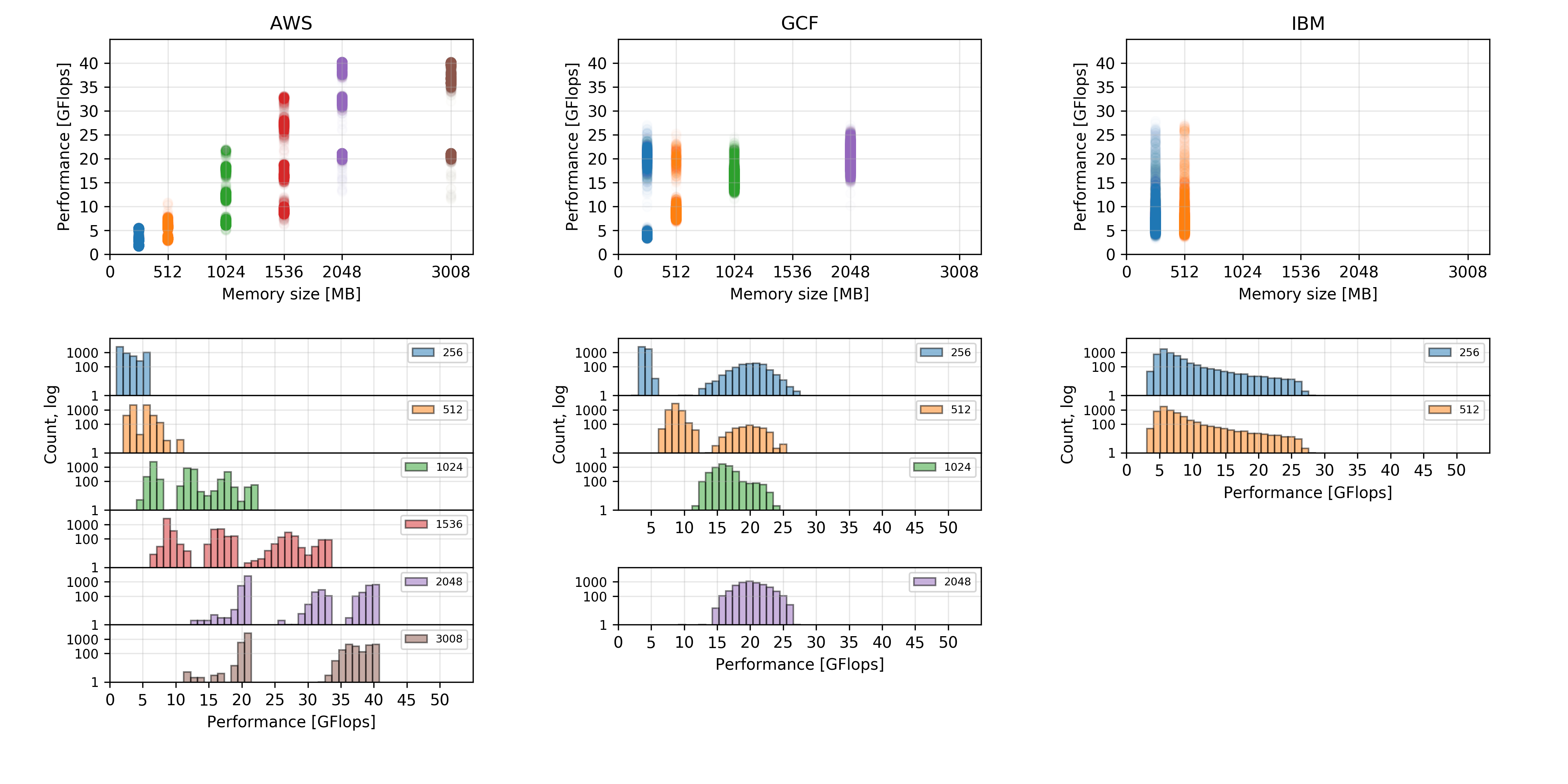}
\caption{Benchmark results for individual service providers.}
\label{fig:score}
\end{figure*}

\section{Objectives}

This paper is a~continuation of work done in \cite{malawski2018performance}, which focused on measurement of cloud function's raw performance. The measurement was justified by the fact that cloud function providers do not supply information on the available performance, or it is done in a~rough and general manner. The main goal of this paper is to provide basis for constructing performance models of scientific workflow applications which are run on FaaS infrastructures. In order to facilitate building such models, we need to have a~more detailed view of the whole process of executing a~cloud function. We determined, that the following areas require more insight:

\begin{itemize}
    \item Performance specific for scientific applications,
    \item Infrastructure provisioning times,
    \item Overhead of the API interface of a~function,
    \item Run time accounting
\end{itemize}

The available raw performance is still important, in this case the measurement was improved by using a~benchmark mimicking the behaviour of scientific software, which mostly relies on vectorized floating point operations. 

Infrastructure provisioning was tested by executing a~benchmark in a~form of a~workflow, which allowed to observe behaviour of FaaS when we start multiple tasks at once. We decided that a~simple {\em bag-of-tasks} type workflow will be the most appropriate for experiments, as it represents a~worst case scenario for the infrastructure provider. The provider needs to find appropriate resources and supply environments for multiple tasks at the same time, hence this allowed for verifying the impact of parallelism limits on workflow execution. 

In the case of overheads, we tried to measure the approximate overhead introduced by using REST API for function calling, this includes message transformation and network delays. 

Finally, the run time accounting was studied by trying to gauge what time is lost by accounting function run time in 100ms increments.

\section{Related Work}

FaaS has been originally designed to host event-based asynchronous applications, coming from Web or mobile usage scenarios. However, there is an~ongoing work on finding other alternative use cases for FaaS, as shown in \cite{baldini2017serverless}, which include exploring areas related to: event processing, API composition and various flow control schemes.
The most promising potential applications related to exploiting the raw processing power include video encoding \cite{Fouladi2017}, where the encoded movie is processed simultaneously in small chunks distributed among many functions.
There are efforts which aim to implement frameworks, like pywren~\cite{jonas2017occupy}, which will allow to perform general purpose computing on FaaS clouds. One of the main features would be to enable dynamic transformation of applications to FaaS model while simultaneously providing deployment services, which would allow for seamless migration to cloud functions. 
In our earlier work~\cite{Malawski2017} we  proposed means to adapt scientific workflows to FaaS, using HyperFlow workflow engine, AWS Lambda and Google Cloud Functions. Another example is \cite{spillner2017faaster}, which describes attempts of porting HPC workloads to FaaS.

Due to the fact, that FaaS infrastructures are a~novelty, the environment is subject to rapid changes. Work done in \cite{leeevaluation} describes the current details of FaaS provider offerings, service types, limitations and costs. Presented results include benchmark results similar to work presented in this paper, albeit performed at smaller scale.

\section{Benchmarking framework}

\begin{figure}[h]
\centering
\includegraphics[width=0.7\columnwidth]{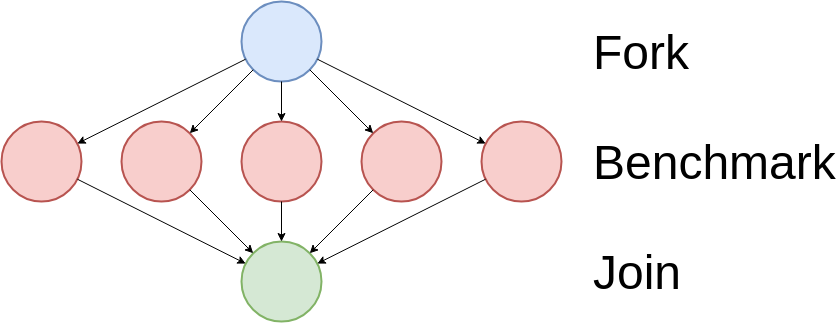}
\caption{Visualization of bag-of-tasks type workflow graph.}
\label{fig:wf}
\end{figure}

The newly developed benchmarking framework is based on framework used in \cite{malawski2017benchmarking}. New features include an~improved testing load and more detailed tracing, which allow us to analyze the complete life cycle of a~function call.
The load used for benchmarking is a~bag-of-tasks workflow, consisting of 5120 tasks, which are expected to start at the same time. \figurename~\ref{fig:wf} depicts a~sample bag-of-tasks type workflow, albeit with only 5 compute tasks for clarity. The large number of tasks was chosen to exceed the concurrency limits imposed by FaaS providers and allows for observing the behaviour of the infrastructure in situations when throttling occurs. HyperFlow \cite{balis2016hyperflow} provided the functionality of workflow management system. HyperFlow was used in our previous work, and has proven to be a~robust and easy to extend by adding support for new infrastructures. Implementing tracing facilities directly into HyperFlow allowed for obtaining an~environment, where the overhead introduced by the system itself was negligible.

Individual tasks contained a~run of Linpack, which is widely accepted benchmark when it comes to measuring floating point performance. The problem size was chosen to be $3408 \times 3408$, which is a~value yielding convenient run time and memory requirements, while providing a~good performance estimate. Supported and tested cloud function providers include: Amazon Cloud Functions (abbr. AWS), Google Cloud Functions (abbr. GCF) and IBM Functions (abbr. IBM).

\section{Results}

The experiments were planned to cover a~broad range of possible memory configurations. Due to the differences in offerings from FaaS providers, each one got assigned a~dedicated list of configurations:

\begin{itemize}
    \item AWS: 256MB, 512MB, 1024MB, 1536MB, 2048MB, 3008MB
    \item GCF: 256MB, 512MB, 1024MB, 2048MB
    \item IBM: 256MB, 512MB
\end{itemize}

The tests were performed from a~node connected to a~network operated by AGH University of Science and Technology located in Krakow, Poland. In the case of FaaS providers, the following regions were used:

\begin{itemize}
    \item AWS: eu-west-1 (Ireland),
    \item GCF: us-central1 (Iowa),
    \item IBM: United Kingdom.
\end{itemize}

\begin{figure}[h]
\centering
\includegraphics[width=\columnwidth]{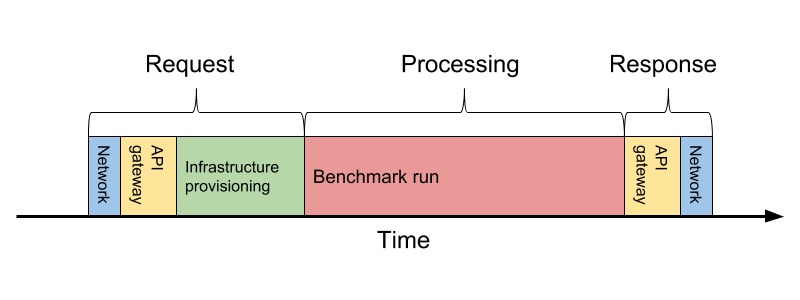}
\caption{Stages of task processing.}
\label{fig:task_stages}
\end{figure}

In order to clarify the naming we divided the time, taken for task processing, into three stages. \figurename~\ref{fig:task_stages} depicts the stages and their respective order.

Depicted task processing time includes the time between the moment when client sends a~function invocation request and time when a~response is received. The first stage is called the {\em Request} stage and includes time spent on network transfer, request processing done by the API handler and infrastructure provisioning. Second stage, {\em Processing} stage consists solely of executing the benchmark. The third stage is called {\em Response} stage and includes creating the response message and network transfer of the response.

\subsection{Computing performance}
\label{ssec:performance}

\begin{figure*}[htb]
\centering
\includegraphics[width=\textwidth]{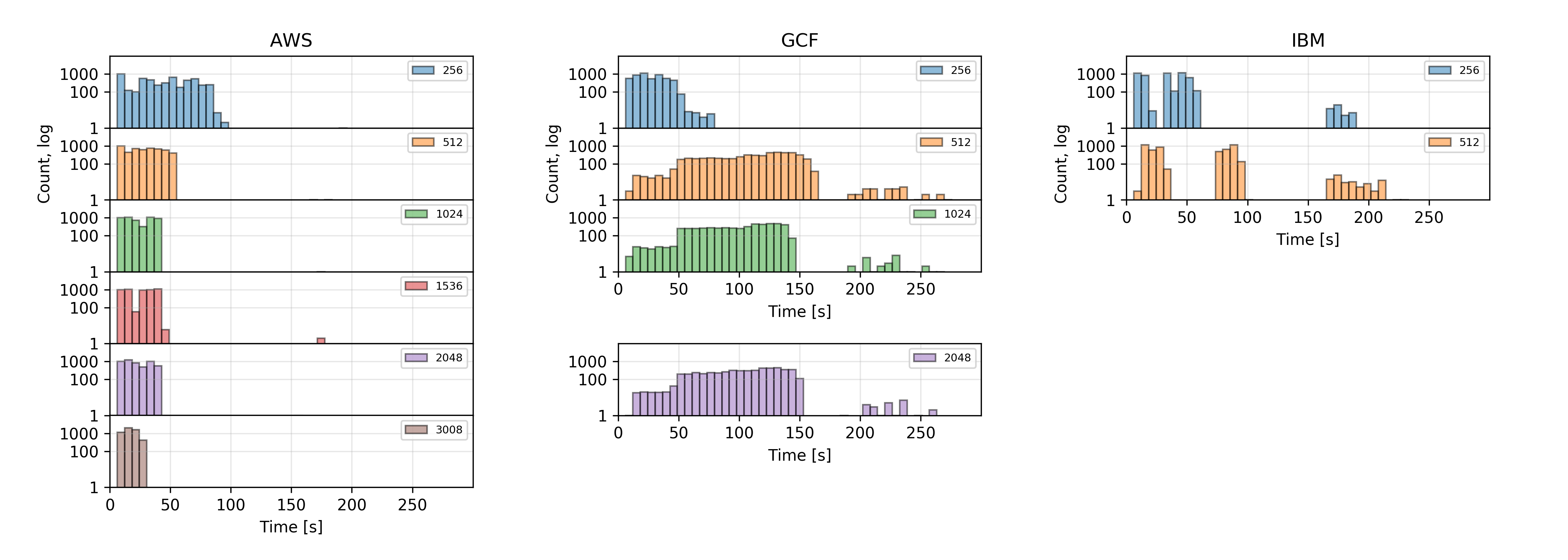}
\caption{Histograms of delay experienced during task start.}
\label{fig:delay}
\end{figure*}

\begin{figure*}[htbp]
\centering
\includegraphics[width=\textwidth]{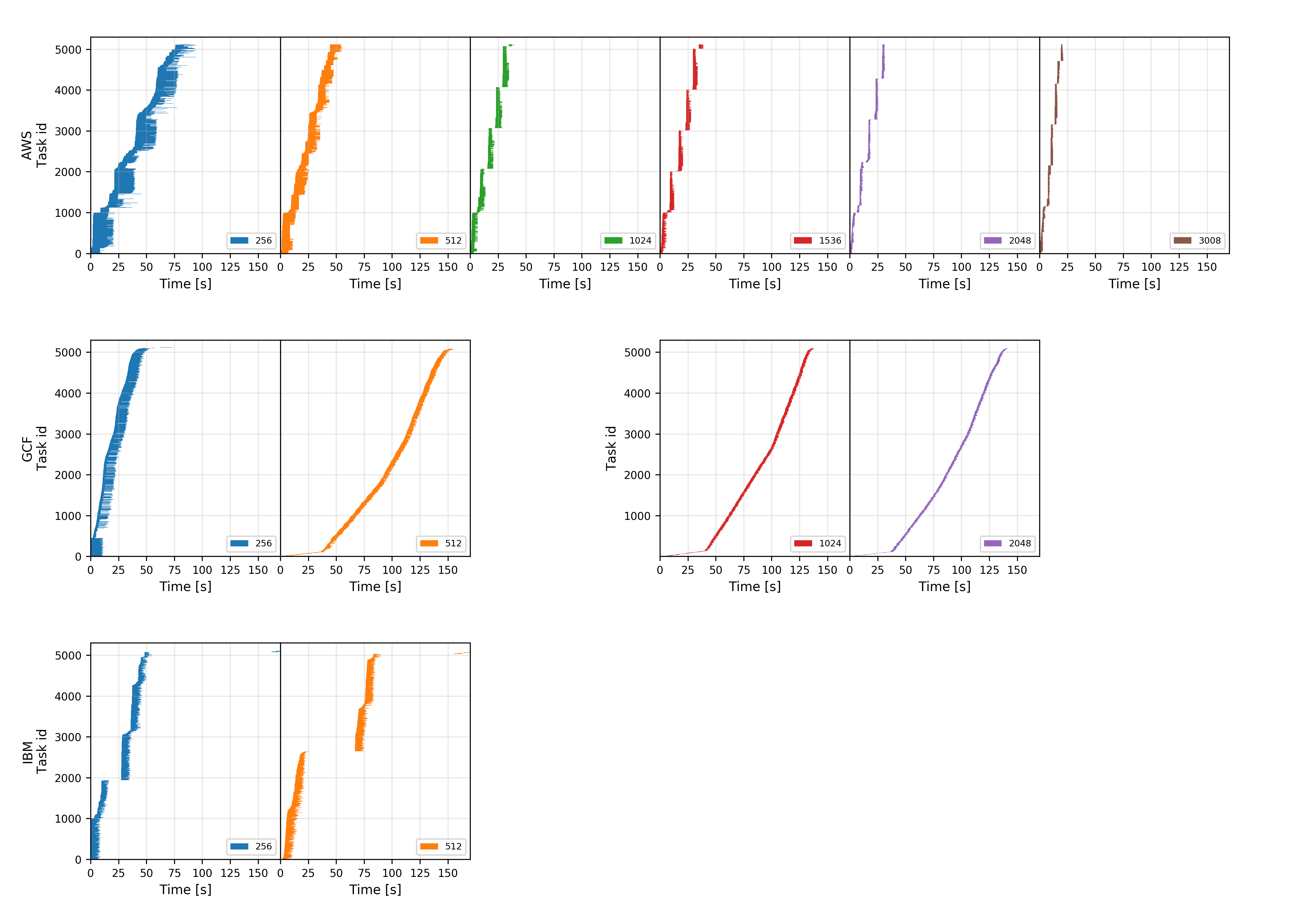}
\caption{Gantt charts of workflow execution.}
\label{fig:gant}
\end{figure*}

\begin{figure*}[htbp]
\centering
\includegraphics[width=\textwidth]{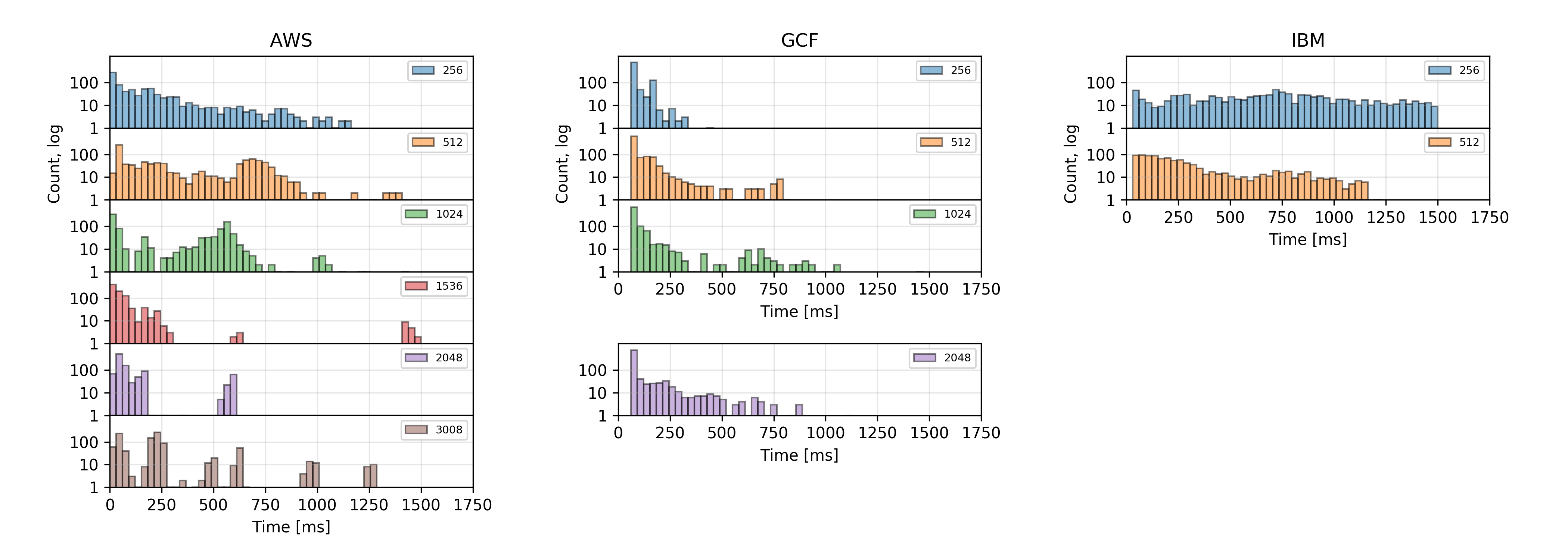}
\caption{API overhead histograms.}
\label{fig:overhead}
\end{figure*}

\figurename~\ref{fig:score} presents results obtained from Linpack benchmarks run as the bag-of-tasks workflow. Results for individual providers are depicted in a~separate column. The upper part of the column contains a~scatter chart visualizing performance relative to declared memory, each value is represented as a~dot. High concentrations of dots (values) result in a~more opaque color. The lower part of the column contains histograms representing the count of sampled performance values for individual memory configurations.

Performance results have been presented in numerical form in table \ref{tab:score}.

\begin{table}[h]
\caption{Performance results}
\begin{center}
\begin{tabular}{|c|c|r|r|}
\hline
\textbf{Provider} &
\begin{tabular}{@{}c@{}}\textbf{Memory size} \\ \textbf{[MB]}\end{tabular} &
\begin{tabular}{@{}c@{}}\textbf{Average} \\ \textbf{performance} \\ \textbf{[GFlops]}\end{tabular} &
\begin{tabular}{@{}c@{}}\textbf{Standard} \\ \textbf{deviation} \\ \textbf{[GFlops]}\end{tabular} \\
\hline
\multirow{6}{*}{AWS} & $256$ & $2.95$ &  $1.38$ \\
& $512$ & $4.62$ &  $1.40$\\
& $1024$ & $10.10$ &  $4.27$\\
& $1536$ & $14.04$ &  $7.18$\\
& $2048$ & $27.26$ &  $8.37$\\
& $3008$ & $27.05$ &  $8.40$\\
\hline
\multirow{4}{*}{GCF} & $256$ & $6.92$ &  $6.23$ \\
& $512$ & $9.54$ &  $3.03$\\
& $1024$ & $16.11$ &  $1.60$\\
& $2048$ & $20.23$ &  $2.03$\\
\hline
\multirow{2}{*}{IBM} & $256$ & $7.35$ &  $3.47$ \\
& $512$ & $7.15$ &  $3.50$\\
\hline
\end{tabular}
\label{tab:score}
\end{center}
\end{table}

In the case of AWS, we can observe obvious relation between memory size and function performance. The larger amount of memory translates to higher average performance. Closer inspection of histograms reveals that the values tend to create multiple clusters of scores, with exception of the 256MB configuration. This might be explained with provisioning infrastructure with different hardware resources or load characteristics. Nonetheless, all the results have one cluster with significantly larger amount of samples than in other clusters. Scores for 2048 MB and 3008 MB configurations are quite similar. The 3008 MB configuration is a~relatively new offering, and it might be possible that the 2048MB configuration (which was the previous maximum value) already offers maximal CPU quota.

Scores obtained from GCF experience similar clustering behaviour, with the exception of the largest configuration. Again, we might be dealing with resources assigned from different pools, while the maximum performance is explicitly limited to some value.

The IBM results present similar performance regardless of the memory configuration. It is worth to notice that the distribution of values is significantly different, than with other providers. The clusters have a~long ,,tail'' in direction of higher values, this might suggest that the functions are able to opportunistically use more resources than expected.

\subsection{Infrastructure provisioning}

The testing workflow is very sensitive to efficiency of infrastructure provisioning. In an~ideal scenario, all tasks start instantly and finish after the same time, but in real-world case we experience various delays, where each individual delayed task is capable of drawing out the total workflow run time. In case of the {\em Request} stage, the total delay (called in this section just ,,delay'' for brevity) is composed of network latency, message processing and infrastructure provisioning. In this case, the majority of the delay comes from infrastructure provisioning and enforcement of parallelism limits.

\figurename~\ref{fig:delay} presents delays, which were encountered during function start. Histograms allow for obtaining an~overview of what can be expect from individual providers. Gantt charts presented in \figurename~\ref{fig:gant} allow for more detailed view of the sequence of task execution. Delay is represented on Gantt charts as the space between vertical axis and start of the task. For readability, providers are split row wise, where each row contains sequence of results for different memory configurations. Inside a~single chart, each task is represented by a~vertical bar spanning from start to end time of computation. Tasks were sorted based on start time for convenience.

In case of AWS we can observe, that larger memory size directly leads to lower delay values. This suggests, that the computing performance is the limiting factor, and infrastructure provisioning occurs in a~fixed, short time. The ,,steps'' visible in Gantt charts are most probably an~effect of parallelism limit which, according to documentation is set to $1000$.

The GCF results show that the start delays are constant across the configurations, with the exception of the 256MB, where tasks experienced the lowest delay. In the Gantt charts we can observe a~pattern for configurations larger than 256MB, where the slope of tasks starts with a~low incline, and rapidly rises between $25$ and $50$ seconds. In contrast to AWS, we don't see steps on Gantt charts, which suggests that function execution is moderated based on rate rather than a~specific level of parallelism.

Delays experienced on IBM suggest, that allocating resources is more time consuming when requesting a~larger amount of memory. Inspecting Gantt charts reveals irregular gaps in start slopes. Additionally, we can observe steps, which are in line with the documented limit of parallelism, which is $1000$.

\subsection{API overhead}

In order to measure overhead introduced by the communication components, namely network and API interface, we need to look at traces of response stage. Response stage doesn't include infrastructure provisioning so it is a~good candidate to measure what is the time taken for network transforming lambda response to an~API message and network transfer, we can assume that the reverse operation (present during the Request stage) will introduce a~similar overhead. \figurename~\ref{fig:overhead} presents histograms of API overheads measured for each provider. In case of AWS and GCF average values are quite low, but it is worth to note that both providers experience a~,,tail'' of values. Although rare, there is a~number of cases where overhead reached up to $1000ms$. Such occurrences have potential to significantly delay the workflow completion. Again, in case of AWS larger functions tend to experience smaller overhead, and the opposite in case of GCF. IBM experiences even distribution of API overhead, where maximum measurement values are smaller for larger function size.

\begin{figure*}[htb]
\centering
\includegraphics[width=\textwidth]{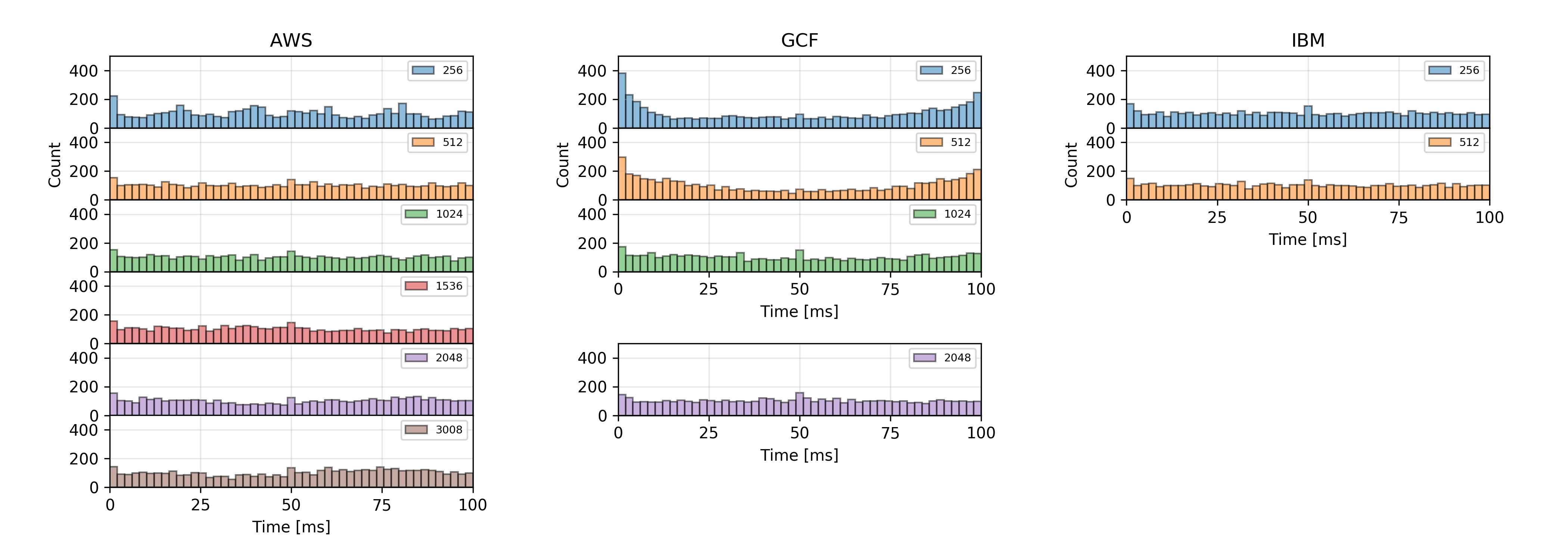}
\caption{Function run time $modulo$ $100ms$ histograms.}
\label{fig:wasted}
\end{figure*}

\subsection{Run time accounting analysis}

The cost of computing on FaaS depends on function's declared memory usage and run time. Memory declaration is done once during the deployment, while run time is measured individually for each function invocation. Conducted experiments showed, that time needed for completing a~task varies even among executions where functions contain identical workloads (as shown in section \ref{ssec:performance}). This might be caused by provisioning environments on different hardware resources and other factors related to {\em noisy neighbours}, where other workloads present on physical machine have an~impact on studied function.

\begin{figure}[h]
\centering
\includegraphics[width=\columnwidth]{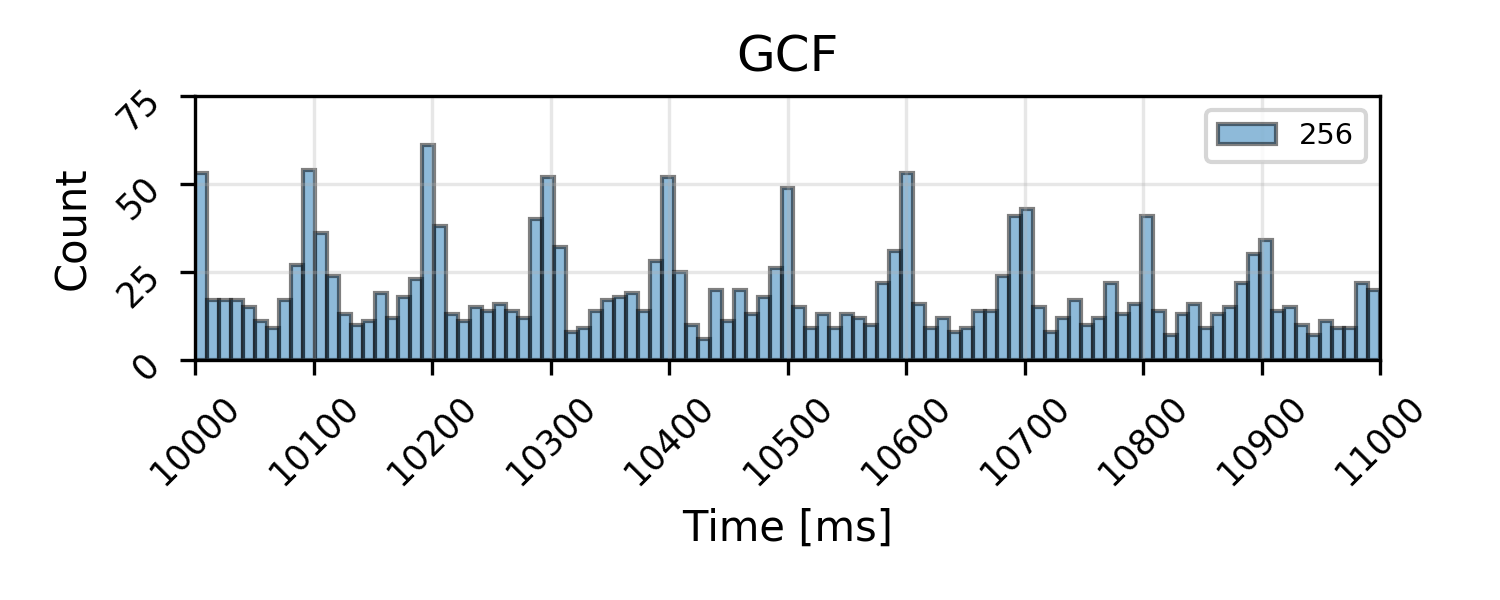}
\caption{Duration histogram for GCF 256MB memory size.}
\label{fig:duration}
\end{figure}

\figurename~\ref{fig:wasted} presents histograms of performing $modulo$ $100$ operation on function run times measured as the difference between actual function start and finish, without any overheads. In the case of AWS and IBM we can see even distribution of values, with slightly higher bars at the start, and in the middle, this might be an~artifact introduced by measuring time with millisecond precision on systems experiencing high load. The GCF case is more interesting, we can see that in low memory configurations, with limited CPU quota, run times tend to end with values close to multiplies of $100ms$. Further investigation resulted in histogram presented in \figurename~\ref{fig:duration}, which depicts distribution of raw task run times, for tasks with run time between $10000ms$ and $11000ms$ for 256MB memory configuration.

We can observe spikes around values which are multiples of $100$. Although it is difficult to explain this phenomena, we can suspect that this is a~side effect of system which is responsible for enforcing the CPU quota. Such system might be dynamically lowering and rising function priority, in order to meet the requested parameters in a~specific time window and thus the available performance may vary with time. Introduced adjustments might be responsible for higher probability of ,,faster system'' near moments in time which are multiples of $100ms$.

\section{Conclusions and future work}

Presented work depicts important characteristics of FaaS computing resources, with the focus on running scientific workflows. Obtained characteristics can provide arguments in discussion, if it is feasible to port certain workflows to FaaS.

The results of the basic performance benchmarks are in line with ones obtained in \cite{malawski2017benchmarking}. Obtained results confirm, that enlarging the memory declaration provides more computing power in case of AWS and GFC, while not providing significant difference in case of IBM. While the memory-performance connection is clearly visible it is worth to note that increasing functions size to values near the maximum value give diminishing gains. Relative performance, obtained on the last step, is not as large as one obtained on previous steps. All results show symptoms of significant variety across individual function invocations.

The REST API overhead proved to be a~significant contribution to the overall time spent on function call, especially for workflows with short tasks (i.e. 500ms of run time). Again significant amount of variety seems to be present in results. Infrastructure provisioning study confirmed that function invocations are subject to limitations. In case of AWS there is clear parallelism limit, while GCF seems to be limited by rate, while IBM results seem to be limited by a~number of invocations per certain time frame. Run time accounting analysis, while very basic, revealed unexpected phenomena encountered on GCF. Tendency, for the workload, to finish in times close to multiplies of $100ms$ suggests that CPU time available to functions is not uniformly distributed through run time.

Future work includes constructing a~performance model for a~set of workflows based on obtained results. Such models, after validation, should result in significant improvements of results obtained from existing and newly developed scheduling algorithms.

\section*{Acknowledgment}

This work was supported by the National Science Centre, Poland, grant 2016/21/B/ST6/01497.

\bibliographystyle{IEEEtran}
\bibliography{IEEEabrv,pcwfc}

\end{document}